\documentclass[aps,amsmath,amssymb,prd,showpacs,floatfix,preprint,superscriptaddress,nofootinbib,12pt]{article}
\pdfoutput=1
\usepackage{jheppub}
\usepackage[utf8]{inputenc}
\usepackage{mathtools}
\usepackage{ragged2e}
\usepackage{emerald}
\usepackage[T1]{fontenc}
\usepackage{bm}
\usepackage{txfonts}
\usepackage{tikz}
\usepackage{ulem}
\usepackage{amsmath,epsfig}
\usepackage{amssymb,amsfonts}
\usepackage{latexsym}

\usepackage{tocvsec2}
\usepackage{subeqnarray}
\usepackage{xcolor}
\let\normalcolor\relax
\usepackage{graphicx}
\usepackage{longtable}

\relax
\def\be{\begin{equation}}
\def\ee{\end{equation}}
\def\bea{\begin{eqnarray}}
\def\eea{\end{eqnarray}}
\newcommand\fverb{\setbox\pippobox=\hbox\bgroup\verb}
\newcommand\fverbdo{\egroup\medskip\noindent%
                        \fbox{\unhbox\pippobox}\ }
\newcommand\fverbit{\egroup\item[\fbox{\unhbox\pippobox}]}

\newcommand{\bear}{\begin{eqnarray}}

\newcommand{\eear}{\end{eqnarray}}

\newcommand{\bsea}{\begin{subeqnarray}}
\newcommand{\esea}{\end{subeqnarray}}
\newbox\pippobox

\def\6{\partial}

\newcommand*{\boxcolor}{orange}
\makeatletter
\renewcommand{\boxed}[1]{\textcolor{\boxcolor}{%
\tikz[baseline={([yshift=-1ex]current bounding box.center)}] \node [rectangle, minimum width=2ex,rounded corners,draw] {\normalcolor\m@th$\displaystyle#1$};}}
 \makeatother

\newcommand{\comments}[1]{}
\newcommand{\foot}{\ensuremath{%
  \mathchoice{\includegraphics[height=3ex]{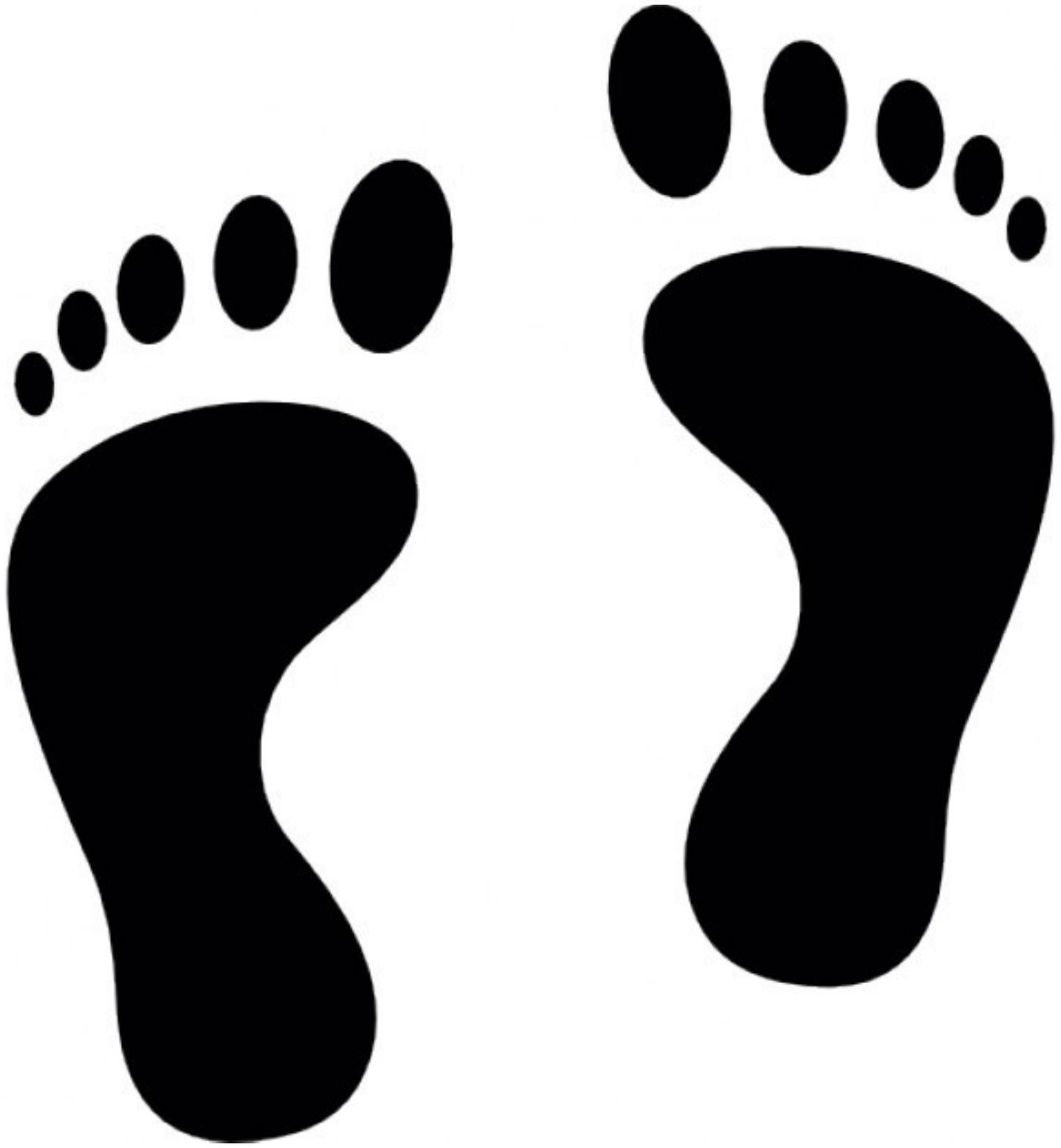}}
    {\includegraphics[height=2.5ex]{foot}}
    {\includegraphics[height=2ex]{foot}}
    {\includegraphics[height=1.5ex]{foot}}
}}

\newcommand{\boxbox}{\ensuremath{%
  \mathchoice{\includegraphics[height=3ex]{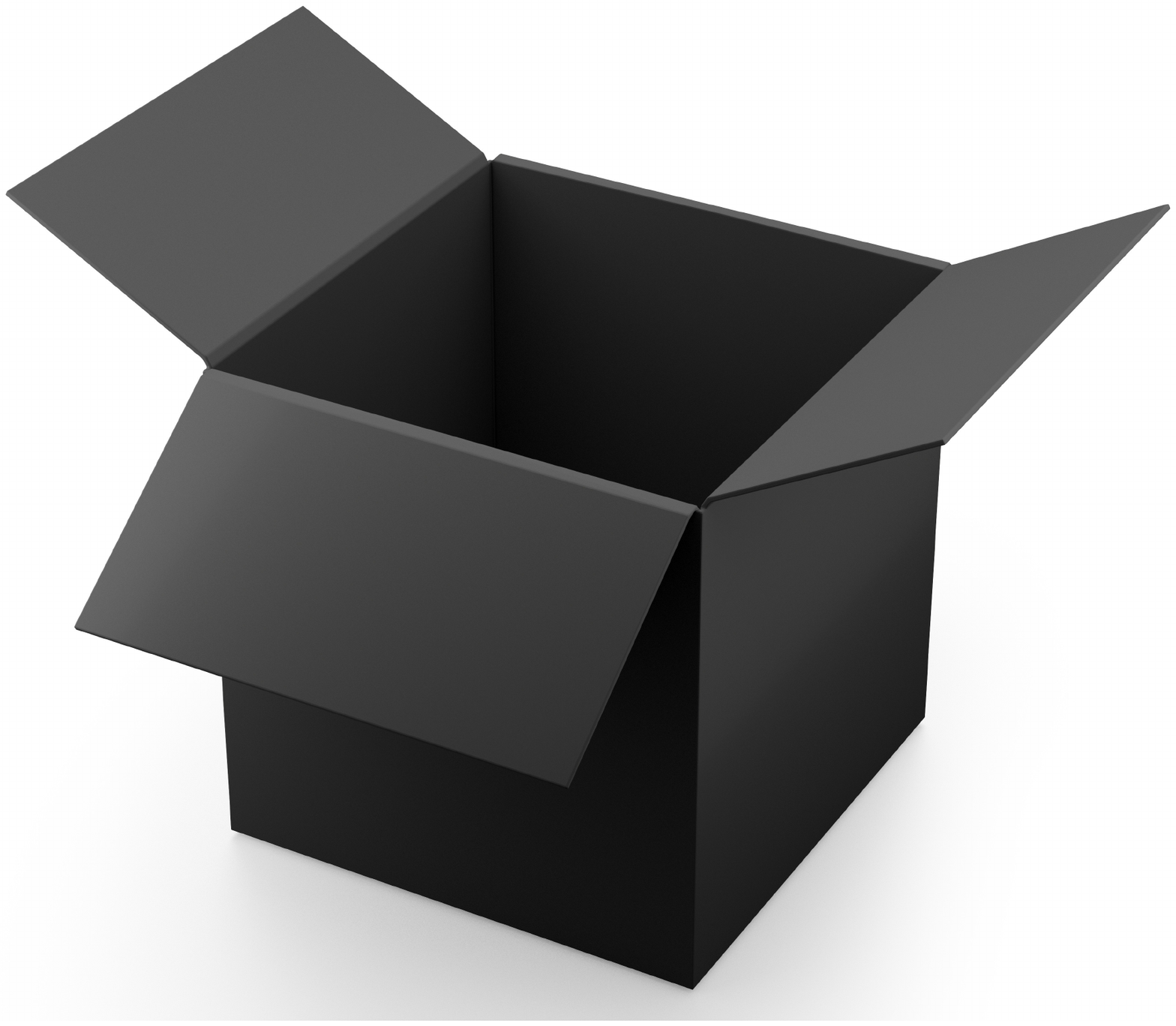}}
    {\includegraphics[height=2.5ex]{boxbox}}
    {\includegraphics[height=2ex]{boxbox}}
    {\includegraphics[height=1.5ex]{boxbox}}
}}

\newcommand{\cool}{\ensuremath{%
  \mathchoice{\includegraphics[height=2ex]{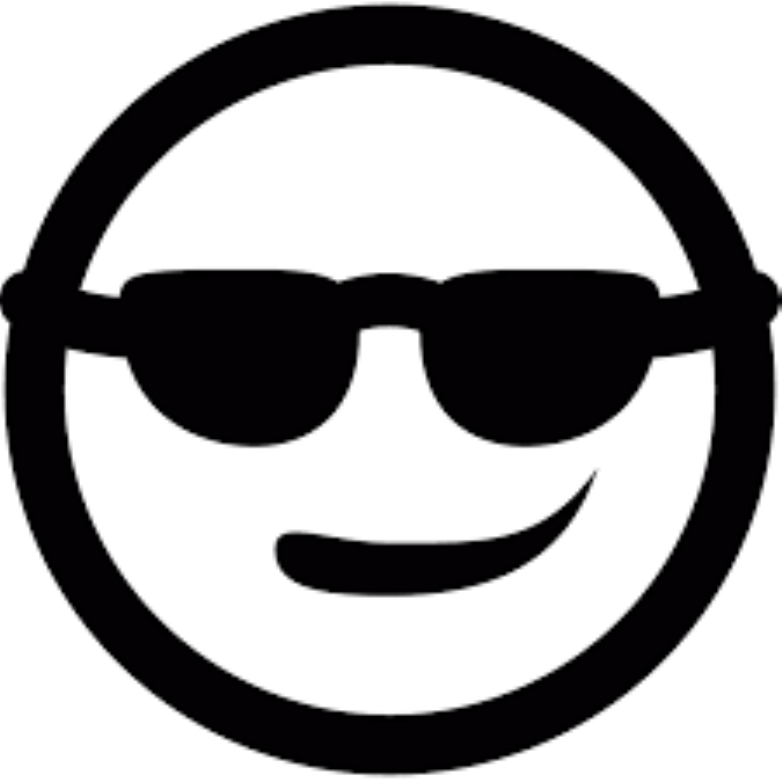}}
    {\includegraphics[height=2ex]{cool}}
    {\includegraphics[height=1.5ex]{cool}}
    {\includegraphics[height=1ex]{cool}}
}}

\DeclareSymbolFont{extraup}{U}{zavm}{m}{n}
\DeclareMathSymbol{\varheart}{\mathalpha}{extraup}{86}
\DeclareMathSymbol{\vardiamond}{\mathalpha}{extraup}{87}

\allowdisplaybreaks[3]

\begin{document}
\subheader{CCTP-2017-2\\ ITCP-IPP 2017/11}
%\begin{flushright}
%\small{CCTP-2016-02\\CCQCN-2016-xx\\SU-ITP-16/XX}
%\end{flushright}

\renewcommand*{\thefootnote}{\fnsymbol{footnote}}
\title{\centering \Huge Diffusivities Bounds and Chaos in Holographic Horndeski Theories}
\author[\, \foot]{Matteo Baggioli}
\affiliation[\foot]{Crete Center for Theoretical Physics, Institute for Theoretical and Computational Physics, Department of Physics, University of Crete, 71003
Heraklion, Greece}
\affiliation[\cool]{Institute of Theoretical Physics, School of Physics, Dalian University of Technology, Dalian 116024, China}
\author[\,\cool]{~\&~~Wei-Jia Li}
\emailAdd{mbaggioli@physics.uoc.gr}
\emailAdd{weijiali@dlut.edu.cn}
\abstract{We study the thermoelectric DC conductivities of Horndeski holographic models with momentum dissipation. We compute the butterfly velocity $v_B$ and we discuss the existence of universal bounds on charge and energy diffusivities in the incoherent limit related to quantum chaos. We find that the Horndeski coupling represents a subleading contribution to the thermoelectric conductivities in the incoherent limit and therefore it does not affect any of the proposed bounds.}

\keywords{}

\maketitle
\renewcommand*{\thefootnote}{\arabic{footnote}}	
\setcounter{footnote}{0}
\section{Introduction}
Strongly correlated materials, \textit{e.g.} strange metals, appear to be characterized by a minimum ''Planckian'' relaxation timescale $\tau^\star \sim\,\hbar/(k_B T)$,  which resonates with the idea of quantum criticality \cite{Sachdev:2011cs}. This fastest possible quantum scale is thought to be responsible for universal transport properties like the linear in T resistivity \cite{Bruin804} and thermal diffusivity \cite{Zhang:2016ofh}. The idea of a minimum timescale, set just by the quantum features of the physics, has already emerged in the past in the context of strongly coupled hydrodynamics \cite{Policastro:2001yc,Kovtun:2004de} and it recently appeared in the framework of quantum chaos \cite{Maldacena:2015waa}. It is generically appealing because it allows to set universal bounds on physical observables which are completely independent of the details of the system. This is exactly what happens with the aforementioned cases:
\begin{equation}
\frac{\eta}{s}\,\geq\,\frac{1}{4\,\pi}\,T\,\tau^\star\,,\qquad \lambda_L\,\leq\,\frac{2\,\pi}{\tau^\star},
\end{equation}
which indeed provide universal bounds for the viscosity to entropy ratio \cite{Policastro:2001yc,Kovtun:2004de} and for the Lyapunov exponent in strongly coupled theories \cite{Maldacena:2015waa}. The approach to this universal regime is tightly connected with the system becoming strongly coupled and with the number of degrees of freedom getting large. Standard perturbative techniques or methods relying on a single particle approximation are not efficient anymore. In the last years the gauge/gravity correspondence has become a valuable tool towards this direction in particular in the realm of condensed matter quantum phases and transport properties \cite{Hartnoll:2016apf}.\\

Recently, using holographic bottom up models, universal bounds on conductivities \cite{Grozdanov:2015qia,Grozdanov:2015djs} have been conjectured :
\begin{align}
&\boxed{\color{black}\sigma\,\geq\,\sigma_{min}}\label{condbound}\\
&\boxed{\color{black}\frac{\kappa}{T}\,\geq\,\mathcal{C}_{min}}\label{heatbound}
\end{align}
where $\sigma$ and $\kappa$ are respectively the electric conductivity and the thermal conductivity at zero current and $\sigma_{min},\mathcal{C}_{min}$ are $\mathcal{O}(1)$ but finite numbers. \color{black} The validity of the proposed inequalities \ref{condbound}, \ref{heatbound} relies on the assumptions that Lorentz invariance is preserved, the boundary theory is $2+1$ dimensional (at least for the bound concerning the electric conductivity) and most importantly that the Maxwell sector is not modified by any direct coupling nor non-linear extensions. In simple words the bulk action must be of the form:
\begin{equation}
\mathcal{L}\,=\,R\,-\,2\,\Lambda\,-\,\frac{1}{4}\,F^2\,+\,\boxbox_{TB}\,+\,\dots \label{simpletheory}
\end{equation}
where $\boxbox_{TB}$ is a generic \textit{disordered} (it has not to be homogeneous) sector which breaks translational invariance and $\dots$ are other eventual matter fields in the model. The bound on the electric conductivity has later been generalized in \cite{Fadafan:2016gmx} for Einstein-Maxwell-Dilaton theories\footnote{See also this recent example \cite{Lucas:2017ggp} which just uses an hydrodynamical approach.}. Deformations to the bulk action \eqref{simpletheory} do eventually violate the bound on the electric conductivity \ref{condbound} but do not affect the validity of the bound on the thermal conductivity \ref{heatbound}. We will discuss this point in more details in the following. \color{black}\\
Inspired by the previous discussion about a possible minimum timescale and the linear in T resistivity feature of the strange metals, S. Hartnoll recently proposed a universal bound for the diffusivities in the incoherent limit \cite{Hartnoll:2014lpa}:
\begin{equation}
D\,\geq\,v^2\,\tau^\star
\end{equation}
where $v$ is a characteristic velocity scale and $D$ the diffusion constants of the system. The bound is conjectured to produce the universal linear in T resistivity observed experimentally in strange metals \cite{Bruin804}. The idea is that in the incoherent limit transport is dominated just by the diffusion of energy and charge and becomes completely insensitive to the microscopic details of the momentum relaxation mechanism leading to a universal quantum behaviour. In absence of a quasiparticle description the velocity $v$, which for common Fermi liquids would be defined by the Fermi velocity $v_F$, has to be thought as the velocity of the collective quantum excitations of the material.\\

Despite some preliminary attempts of verifying such a conjecture using holography \cite{Amoretti:2014ola}, without the precise definition of the velocity scale $v$ the prescription is not practicable and testable. Just recently M. Blake proposed the identification of $v$ with the butterfly velocity $v_B$ \cite{Blake:2016wvh,Blake:2016sud} and made the Hartnoll bound operative in arbitrary chaotic systems.\\
The butterfly velocity \color{black} $v_B$ \color{black} measures the speed of propagation of information through a quantum system and it can be generically defined by the out-of-time correlator:
\begin{equation}\label{OTOC}
\langle\,\left[\mathcal{V}(x,t)\,\mathcal{W}(0,0)\right]^2\,\rangle_\beta\,\sim\,e^{\lambda_L\,\left(t-t^*\,-\,|x|/v_B\right)},
\end{equation}
where $\mathcal{V},\mathcal{W}$ are two generic Hermitian operators, $\lambda_L$ is the Lyapunov exponent, $t^*$ is the so called scrambling time and $\beta$ is just the thermal timescale. The previous formula is nothing else but the quantum version of the following statement:
\begin{equation*}
\,\,\,\,\,\,\,\,\,\,\,\,\,\,\,\,\textit{classical chaos}\,\,\equiv\,\,\textit{exponential dependence on the initial conditions,}
\end{equation*}
which is now formulated in terms of four point functions and not two points functions anymore.\\
The butterfly velocity can be computed holographically considering a localized shockwave perturbing the initial spacetime and boosting the backreaction to the background in a later time \cite{Roberts:2014isa,Roberts:2016wdl}; the butterfly velocity $v_B$ simply describes the rate of growth of such an excitation. \\

Identifying, as proposed by Blake, the velocity scale with the butterfly velocity $v_B$ the refined Hartnoll bound takes the following form:
\begin{equation}\label{thebound}
\boxed{\color{black}D\,\geq\,v_B^2\,\frac{\hbar}{k_B\,T}}
\end{equation}
and it can be tested both theoretically and experimentally\footnote{See \cite{Ling:2016wuy,Ling:2016ibq} for some other recent holographic applications regarding the butterfly effect and condensed matter topics.} .\\

In simple holographic models with momentum relaxation the bound \eqref{thebound} appears to be respected \cite{Blake:2016wvh,Blake:2016sud} and to be insensitive to the microscopic details of the theory \cite{Davison:2016ngz} or the presence of a finite charge density \cite{Kim:2017dgz}. Moreover, the conjectured inequality \eqref{thebound} holds in several \textit{non-holographic} models like the SYK model \cite{Gu:2016oyy,Davison:2016ngz}, weakly coupled Fermi-liquids \cite{Aleiner:2016eni}, diffusive metals \cite{Swingle:2016jdj}, critical Fermi surface models \cite{Patel:2016wdy}, Bose-Hubbard models \cite{Bohrdt:2016vhv} and electron-phonon bad metals \cite{2017arXiv170507895W}. It is therefore a very interesting and valuable question to understand to which extent \eqref{thebound} is generic, and if it is why it is that. In the light of the recent developments in this direction it is convenient for the discussion to separate the charge sector from the energy one\footnote{At zero charge density $q=0$, the charge and energy sector are clearly decoupled and the diffusivities matrix becomes diagonal:
\begin{equation}
\mathcal{D}\,=\,\begin{pmatrix} 
D_c\equiv\frac{\sigma}{\chi} & 0 \\
0 & D_e\equiv\frac{\kappa}{c_\rho} 
\end{pmatrix}
\end{equation}
In the incoherent limit this is still generically true because the off diagonal part $D_{ij}$ turns out to be suppressed by $\rho/k$, where $k$ is the momentum dissipation strength.  \color{black}See \cite{Kim:2017dgz} for a concrete example of such suppression.\color{black}}.\\[0.3cm]
%\newpage
%\justify
\textbf{Charge sector}\\[0.1cm]
In the holographic models the DC (zero frequency) electric conductivity takes a generic and simple structure\footnote{Notice that such bipartite structure was suggesting the possibility of reproducing the Cuprates scalings within holographic theories \cite{Blake:2014yla}. Unfortunately we do not know any explicit and succesful model in that direction yet \cite{Amoretti:2016cad}.} \cite{Lucas:2015vna,Lucas:2015lna,Davison:2015bea}:
\begin{equation}\label{du}
\sigma_{DC}\,=\,\sigma^{(inc)}\,+\,\underbrace{\frac{q^2}{\mathcal{M}_h^2}}_{\xrightarrow[\text{inc}]{}\,0}
\end{equation}
where $\mathcal{M}_h^2>0$ is the effective, and model dependent, graviton mass computed at the horizon and $q$ the charge density. The second term in \eqref{du}, resembling the Drude conductivity\footnote{\color{black}This term corresponds to the dissipative contribution only at leading order in the momentum relaxation strength \Big($\mathcal{O}\left(k^{-2}\right)$ in our notations\Big); once one includes higher order terms the situation becomes more complicated and the second term does not correspond anymore to the weight of the drude pole \cite{Davison:2015bea}. We thank Blaise Gouteraux for clarifications about this point.\color{black}}, generically drops to zero in the incoherent limit\footnote{We will define such a limit in details in section \ref{sec3}; for the moment we can think of it just as the limit of very fast momentum relaxation.} whereas the first term, as the name suggests, (at least generically) survives and determines indeed $\sigma_{DC}$ in such a limit.
The bound for the electric conductivity \eqref{condbound}, which basically states that:
\begin{equation}
\sigma^{(inc)}\,>\,0
\end{equation}
was formulated for \textit{simple} holographic models \color{black} of the form \eqref{simpletheory} \color{black}. Nowadays it is clear that such a statement is not generic and can be violated in several ways exploiting higher derivatives couplings between the translations breaking (TB) sector and the charge one \cite{Baggioli:2016oqk,Gouteraux:2016wxj,Baggioli:2016pia,Garcia-Garcia:2016hsd}, using dilatonic couplings \cite{Gouteraux:2014hca,Kiritsis:2015oxa} or modifying the Maxwell term with higher order corrections \cite{Baggioli:2016oju}. In the same way the bound for the charge diffusivity \eqref{thebound} , where we set $D=D_c\equiv \sigma/\chi$, can be violated using the same kind of holographic ``homogeneous''\footnote{By homogeneous we mean models which break translational invariance retaining the metric homogeneous thanks to some global symmetry. To be more precise we are thinking about massive gravity models \cite{Vegh:2013sk,Andrade:2013gsa,Baggioli:2014roa}, Q-lattices \cite{Donos:2013eha} and helical backgrounds \cite{Donos:2012js,Donos:2014oha}.} bottom up models \cite{Baggioli:2016pia} and also exploiting really disordered holographic systems \cite{Lucas:2016yfl}.  Somehow it is not so surprising that this happens, because the charge susceptibility $\chi$, despite all the other quantities, is not given purely in terms of horizon data but it depends of the details of the full geometry (meaning on the full RG flow of the theory). As a consequence one does not necessarily expect any sign of universality in the charge sector; it is indeed quite easy to ``mess up'' with the Maxwell structure in order to go beyond the universal bounds \eqref{condbound}, \eqref{thebound}.\\[0.25cm]
\textbf{Energy sector}\\[0.1cm]
For the energy sector the story is quite different. The heat conductivity and the energy diffusion constant are directly connected to the gravitational sector where it is much harder to introduce well-behaved modifications. In addition, both the thermal conductivity and the heat capacity (and the butterfly velocity) can be defined in terms of horizon data and therefore are more likely to exhibit a universal behaviour. So far, there are no holographic examples of the violation of the thermal conductivity \eqref{heatbound} and energy diffusivity \eqref{thebound} (where $D=D_e=\kappa/c_\rho$) bounds\footnote{It is anyway possible to find out models where $\kappa/T$ is arbitrarily small, but never exactly zero. One example of that is the model introduced in \cite{Baggioli:2014roa} and analyzed later on in \cite{Baggioli:2016pia}.}. Only recently a computation done with a non homogeneous generalization of the SYK  model \cite{Gu:2017ohj} seems to provide a counterexample to:
\begin{equation}
\frac{D_e\,T}{v_B^2}\,\equiv\,\frac{\kappa\,T}{c_\rho\,v_B^2}\,\geq\,\mathcal{C}
\end{equation}
where the sign of the inequality appears to be reversed.\\
Despite the analogy with the charge sector, higher derivatives corrections by themselves do not affect the  heat conductivity by any means \cite{Cheng:2014tya,Baggioli:2016pia}. The target of this paper is to investigate the following question:
\begin{equation*}
\color{black}\boxed{\color{black} \textit{To which extent the bounds }\left\{\frac{\kappa}{T}\,\geq\,\mathcal{C}_0,\,\frac{D_e\,T}{v_B^2}\,\geq\,\mathcal{C}_1\right\}\,\textit{are generic in holographic theories ??}}
\end{equation*}
Inspired by the recent results in the charge sector, we will study these issues in holograhic (healthy) bottom up toy models where the translations breaking sector is directly coupled to the gravity sector itself. \color{black} The idea is to test if higher derivative couplings in the gravitational sector could produce consistent modifications to the diffusion constants and the consequent violation of the universal bounds as it happens in the charge sector \cite{Gouteraux:2016wxj}.\color{black}\\[0.25cm]
\textbf{Horndeski models}\\[0.1cm]
The simplest and easiest way to modify General Relativity are the so called scalar-tensor theories, where one additional scalar and real degree of freedom is introduced violating the strong equivalence principle. The prototype of those models is certainly the Brans-Dicke theory but in the last decades a lot of progress has been done; one example is what takes the name of Horndeski theories \cite{Horndeski:1974wa,Deffayet:2013lga,Sotiriou:2013qea}. They represent simple modifications of the Einstein-Hilbert action defined by a new scalar degree of freedom derivatively coupled to gravity. The easiest example consists in a coupling to the Einstein tensor as follows:
\begin{equation}
\sim\,G^{\mu\nu}\,\partial_\mu \phi\,\partial_\nu \phi\,=\,\left(R^{\mu\nu}\,-\,\frac{1}{2}\,R\,g^{\mu\nu}\right)\,\partial_\mu \phi\,\partial_\nu \phi
\end{equation}
where the scalar $\phi$ enjoys shift symmetry.\\
These theories have received a lot of interest recently in particular for their cosmological properties and their connections with Galileons \cite{Nicolis:2008in} and because they avoid ghost instabilities thanks to the equations of motion which remain $2_{nd}$ order. They also have been analyzed in the context of holography and several black hole solutions, with their corresponding thermodynamics, have been studied \cite{Charmousis:2012dw,Feng:2015oea,Feng:2015wvb}.\\

On the other side the linear Stueckelberg model \cite{Andrade:2013gsa} has been identified as a particularly interesting and easy effective toy model to introduce momentum relaxation into the realm of the AdS-CMT correspondence. The model makes use of a set of massless scalar fields linearly sourced in the spatial directions:
\begin{equation}\label{scalars}
\phi^I\,\sim\,x^I
\end{equation}
which therefore produce momentum relaxation in the dual picture. It clearly represents a massive gravity theory written down in the Stuckelberg formalism where those scalars are indeed the Goldstone modes for the broken translational invariance. Massive gravity, in all its forms \cite{Baggioli:2014roa,Alberte:2015isw,Vegh:2013sk,Blake:2013owa} stands like the universal effective holographic theory for momentum relaxation, \color{black}where the relaxation rate is indeed defined by \cite{Davison:2013jba}:
\begin{equation}\label{TAU}
{\tau_{rel}}^{-1}\,=\,\frac{s}{2\,\pi\,(\epsilon+P)}\mathcal{M}_h^2
\end{equation}
where $\mathcal{M}_h^2$ is the effective graviton mass\footnote{To be more precise, it is the mass of the helicity 1 component of the graviton.} computed at the position of the event horizon $r=r_h$, $s$ the entropy density and $\epsilon,\,P$ respectively the energy density and the pressure of the dual CFT. \color{black} Studying the phenomenology of these holographic theories is particularly simple because of the possibility of getting all the DC conductivities analytically \cite{Donos:2014cya,Amoretti:2014mma}.\\[0.2cm]
In this paper we study the thermoelectric conductivities in a recently introduced holography Horndeski model \cite{Jiang:2017imk}\footnote{Horndeski theories have been applied earlier in the context of holographic superconductors in \cite{Kuang:2016edj}.} where the Stuckelberg fields are derivatively coupled to the Einstein tensor, \textit{i.e.} $G^{\mu\nu}=R^{\mu\nu}-\frac{1}{2}\,g^{\mu\nu}R$, as follows:
\begin{equation}
\sim\,-\frac{1}{2}\,\left(g^{\mu\nu}\,-\,\gamma\,G^{\mu\nu}\right)\,\partial_\mu \phi^I\,\partial_\nu \phi^I\,,\qquad \phi^I\,\sim\,x^I
\end{equation}
where $\gamma$ is the \textit{new} Horndeski coupling.\\
We study the validity of the proposed bounds on thermal conductivity and energy diffusion and we find that they both hold because the Horndeski deformation $\gamma$ represents a subleading contributions to the thermoelectric conductivities in the incoherent limit and therefore it does not modify to any extent the analysis.\\[0.4cm]
The manuscript is organized as follows: in section \ref{sec1} we define the model and the black brane background solution we will work with; in section \ref{sec2} we discuss the thermoelectric transport properties of the dual CFT and the possible existence of bounds related to them; in section \ref{sec3} we investigate the recently proposed bounds on diffusivities related with quantum chaos in our framework; in section \ref{sec4} we summarize our results and we propose future directions; finally in the appendices \ref{appen}, \ref{app1} and \ref{app2} we provide the reader with more details about the computations appearing in the main text.
\section{The model}\label{sec1}
We consider an extension of the linear Stueckelbergs model \cite{Andrade:2013gsa} recently introduced in \cite{Jiang:2017imk}\footnote{In order to simplify the model we slightly changed the notations and we set the magnetic field B to zero. One can recover the results of \cite{Jiang:2017imk} fixing there $\alpha=\kappa=1$, $\lambda=k$ and $B=0$.}. \color{black}The model and similar solutions have been also discussed in \cite{Anabalon:2013oea,Cisterna:2014nua}. \color{black} More in details we consider the following $3+1$ dimensional model:
\begin{equation}\label{themodel}
\mathcal{S}\,=\,\int\,d^4x\,\sqrt{-g}\,\left(\,R\,-\,2\,\Lambda\,-\,\frac{1}{4}\,F^2\,-\,\frac{1}{2}\,\left(\,g^{\mu\nu}\,-\,\gamma\,G^{\mu\nu}\right)\,\sum_{i=1}^2\,\partial_\mu \phi^i\,\partial_\nu\phi^i\,\right)
\end{equation}
where $G^{\mu\nu}=R^{\mu\nu}-\frac{1}{2}\,R\,g^{\mu\nu}$ is the usual Einstein tensor. We fix $16\,\pi\,G_N=1$.\\
The parameter $\gamma$ is the Horndeski coupling and it is the new ingredient in the story; setting $\gamma=0$ one recovers the results of \cite{Andrade:2013gsa}.\\
The generic equations of motion for the system are:
\begin{align}
& G_{\mu\nu}\,+\,\Lambda\,g_{\mu\nu}\,-\,\frac{1}{2}F_{\mu\rho}{F^{\rho}}_\nu\,+\,\frac{1}{8}\,F^2\,g_{\mu\nu}\,-\,\frac{1}{2}\,\sum_{i=1}^2\,\left(\partial_\mu \,\phi^i\,\partial_\nu \,\phi^i\,-\,\frac{1}{2}\,g_{\mu\nu}\,\left(\partial \phi^i\right)^2\right)\nonumber\\
&-\,\sum_{i=1}^2\,\frac{\gamma}{2}\,\Big\{\frac{1}{2}\,\partial_\mu \,\phi^i\,\partial_\nu\, \phi^i\,R\,-\,2\,\partial_\rho\, \phi^i\,{\partial_{(\mu}\,\phi^i\,R_{\nu)}}^\rho\,-\,\partial_\rho\,\phi^i\,\partial_\sigma\,\phi^i\,{{{R_\mu}^\rho}_\nu}^\sigma\nonumber\\
&-\,\left(\nabla_\mu\,\nabla^\rho\,\phi^i\right)\,\left(\nabla_\nu\,\nabla_\rho\,\phi^i\right)\,+\,\left(\nabla_\mu\,\nabla_\nu\,\phi^i\right)\,\Box\,\phi^i\,+\,\frac{1}{2}\,G_{\mu\nu}\,\left(\partial\,\phi^i\right)^2\nonumber\\
&-\,g_{\mu\nu}\,\left[-\,\frac{1}{2}\,\left(\nabla^\rho\,\nabla^\sigma\,\phi^i\right)\,\left(\nabla_\rho\,\nabla_\sigma\,\phi^i\right)\,+\,\frac{1}{2}\,\left(\Box \phi^i\right)^2\,-\,\partial_\rho\,\phi^i\,\partial_\sigma\,\phi^i\,R^{\rho\sigma}\right]\Big\}\,=\,0\\[0.2cm]
&\nabla_\mu\,\left[\left(g^{\mu\nu}\,-\,\gamma\,G^{\mu\nu}\right)\,\partial_\nu\,\phi^i\,\right]\,=\,0\,\qquad \nabla_\nu\,F^{\nu\mu}\,=\,0
\end{align}
where $\Box=\frac{1}{\sqrt{g}}\partial_\mu\left(\sqrt{g}g^{\mu\nu}\partial_\nu\right)$ is the Laplace-Beltrami operator.\\
In order to avoid ghosty excitations in the $\phi$ sector (see \cite{Jiang:2017imk}) we have to restrict the Horndeski coupling to the range:
\begin{equation}\label{cons}
-\,\infty\,<\,\gamma\,\leq\,-\,\frac{1}{\Lambda}
\end{equation}
where $\Lambda<0$.\\
We look for black hole (BH) solutions with asymptotic AdS spacetime of the form:
\begin{align}\label{ansatz}
&ds^2\,=\,-\,h(r)\,dt^2\,+\,\frac{dr^2}{f(r)}\,+\,r^2\,dx^i\,dx^i\nonumber\\
&A\,=\,A_t(r)\,dt\,,\qquad \phi^i\,=\,k\,x^i
\end{align}
The corresponding equations of motion become:
\begin{align}
&r^4\, {A_t'}^2+2\,h \left(\frac{k^2\, r^2+2 \,\Lambda\,  r^4}{f}+\gamma \, k^2+2 \, r^2\right)+4 \,  r^3 \,h'\,=\,0\,,\\
&r^4\, f\, {A_t'}^2+2 \,h\, \left(r^2 \left(2  \, r\, f'\,+\, k^2\,+\,2\,  \Lambda\,  r^2\right)+f \left(2 \, r^2-\gamma\, 
   k^2\right)\right)\,=\,0\,,\\
   &h\,\left\{f\, \left[\, r^4\, {A_t'}^2-2 \,  r^4 \,h''-\,r\, h'\,\left(\gamma  \,k^2+2 \,  r^2\right)\,\right]-\,  r^4\, f'\,
   h'\right\}\nonumber\\&-h^2 \left(r \,f'\, \left(\gamma  \,k^2+2 \,  r^2\right)-2\, \gamma  \,k^2 \,f+4 \,  \Lambda\,  r^4\right)+ \, r^4\, f\, {h'}^2\,=\,0\,,\\
   &2 \,r \,f\, h\, A_t''\,+\,A_t'\, \left(r\, h\, f'\,+\,f\, \left(4 \,h\,-\,r\, h'\right)\right)\,=\,0\,.
\end{align}
The BH solution can be easily obtained and it reads:
\begin{align}
&h(r)\,=\,U(r)\,f(r)\,,\qquad U(r)\,=\,e^{\frac{\gamma \, k^2}{2 \,  r^2}}\,,\\
&A_t(r)\,=\,\mu \,-\,\frac{\sqrt{\pi }  \,q\, \text{erfi}\left(\frac{\sqrt{\gamma } \,k }{2\, r}\right)}{\sqrt{\gamma } \,k }\\
&f(r)\,=\,-\frac{e^{-\frac{\gamma  k^2}{4 r^2}} \left(2 \sqrt{\gamma } k  r e^{\frac{\gamma  k^2}{4r^2}}
   \left(k^2 (3+\gamma  \Lambda )+2 \Lambda  r^2\right)-\sqrt{\pi }\, \text{erfi}\left(\frac{\sqrt{\gamma } k }{2
    r}\right) \left(\gamma  k ^4 (3 +\gamma  \Lambda )+3 q^2\right)\right)}{12 \sqrt{\gamma }  k
    r}-\frac{f_0\, e^{-\frac{\gamma  k^2}{4 r^2}}}{r}
\end{align}
where $\text{erfi}(\mathcal{y})=-i\,\text{erf}(i\,\mathcal{y})$ is the immaginary error function\footnote{The error function is defined as: \begin{equation}
\text{erf}(q)\,=\,\frac{1}{\sqrt{\pi}}\,\int_{-q}^q\,e^{-t^2}\,dt
\end{equation}}. Setting $\gamma=0$\footnote{Notice that erfi$(x)\sim\frac{2\,x}{\sqrt{\pi}}$ in the limit of small argument $x\rightarrow 0$.} we recover the results of \cite{Andrade:2013gsa}.
The integration constant $f_0$ is fixed by the requirement of having a BH horizon located at $r=r_h$ which corresponds to impose $f(r_h)=0$. We also define $\mu,q$ the chemical potential and the charge density of the dual CFT, which turn out to be related as follows \color{black}:
\begin{equation}
q\,=\,\mu\,\frac{\sqrt{\gamma } \,k}{\sqrt{\pi } \,\text{erfi}\left(\frac{\sqrt{\gamma } \,k}{2 \,r_h}\right)}\label{EE3}
\end{equation}
where this last relation is obtained by requiring the regularity of the gauge field $A_\mu$ at the horizon, \textit{i.e.} $A_t(r_h)=0$.\color{black}\\
The temperature of the BH is obtained as:
\begin{equation}\label{Tdef}
T\,=\,\frac{\sqrt{f'(r_h)\,h'(r_h)}}{4\,\pi}\,=\,-\frac{e^{\frac{\gamma \, k^2}{4 \,r_h^2}} \left( q^2+4\, \Lambda\,  r_h^4+2  \,k^2
   \,r_h^2\right)}{16 \,\pi\,r_h^3}
\end{equation}
The entropy density is defined by:
\begin{equation}\label{entropydens}
s\,=\,4\,\pi\,r_h^2\,\left(1\,-\,\frac{\gamma}{2\,r_h^2}\,k^2\right)
\end{equation}
The details of its derivation are given in appendix \ref{appen}.\\
It is quite interesting to notice that the requirement of having positive entropy density corresponds to the no ghosts condition \eqref{cons} found in \cite{Jiang:2017imk}\footnote{Notice that in the large $k$ limit we have:
\begin{equation}
r_h\,\sim\,\frac{k}{\sqrt{-2\Lambda}}\,\quad \rightarrow \quad\,1\,-\,\frac{\gamma}{2\,r_h^2}\,k^2\,=\,1\,+\,\gamma\,\Lambda
\end{equation}}.
\section{Thermoelectric DC conductivities}\label{sec2}
The response of the system to a small electric field E and a small temperature gradient $\nabla T$ is encoded in the matrix of the thermoelectric conductivities which can be defined as:
\begin{equation}
\left(
    \begin{array}{c}
      \mathcal{J} \\
      \mathcal{J}^Q
    \end{array}
  \right)\,=\,\left(
    \begin{array}{cc}
      \sigma & \alpha\,T\\
      \bar{\alpha}\,T & \bar{\kappa}\,T
    \end{array}
  \right)\left(
    \begin{array}{c}
      E \\
      -\frac{\nabla T}{T}
    \end{array}
  \right)
\end{equation}
where $\mathcal{J},\, \mathcal{J}^Q$ are the electric and heat currents and $\sigma,\,\alpha=\bar{\alpha}\,,\bar{\kappa}$\footnote{Whenever time reversal is preserved we have $\alpha=\bar{\alpha}$ because of the Onsager relations (see for example \cite{Donos:2017mhp}).} are respectively the electric, thermoelectric and thermal conductivities.\\
Using the method of \cite{Donos:2014cya}\footnote{\color{black}See also \cite{Amoretti:2014mma,Blake:2015ina,Amoretti:2015gna,Amoretti:2017xto,Gouteraux:2014hca,Lucas:2015lna,Lucas:2015pxa,Lucas:2015vna,Kim:2014bza,Ge:2014aza,Kim:2015wba} for generalizations and subsequent works on the topic.\color{black}} one can obtain the full set of thermoelectric DC conductivities of the dual (deformed) CFT, which takes the following form:
\begin{align}\label{DC}
&\sigma\,=\,1\,+\,\frac{q^2}{\mathcal{M}_h^2}\,,\qquad\alpha=\bar{\alpha}=\,\frac{4\, \pi\,r_h^2\,q}{\mathcal{M}_h^2}\,,\qquad\bar{\kappa}\,=\,\frac{16 \,\pi ^2 \,T\,r_h^4}{\mathcal{M}_h^2}\,,\qquad\kappa\,=\,\frac{16 \,\pi ^2 \,T\,r_h^4}{\mathcal{M}_h^2\,+\,q^2}
\end{align}
where $\kappa=\bar{\kappa}\,-\,\bar{\alpha}\,\alpha\,T/\sigma\,$ represents the thermal conductivity at zero electric current. The details about the derivation of the transport matrix are given in appendix \ref{app1}.\\
$\mathcal{M}_h^2$ is the effective graviton mass\footnote{To be precise it corresponds to the effective mass of the vectorial part of the metric. A priori, in Lorentz violating massive gravity theories, the latter can be different from the mass of the helicity 2 component of the metric; see \cite{Alberte:2015isw,Alberte:2016xja}.} computed at the horizon:
\begin{equation}\label{effmass}
\mathcal{M}_h^2\,=\,k^2\,r_h^2\,-\,\gamma\,\left(4\,\pi\,k^2\,r_h\,T\,e^{-\frac{\gamma  \,k^2}{4\,r_h^2}}\right)
\end{equation}
\color{black} At low temperature the Horndeski coupling represents a subleading contribution to the graviton mass; on the contrary at high temperature, \textit{i.e.} $r_h\sim T$, the new term scales exactly as the one of the linear theory. \color{black}
We notice from \eqref{effmass} that at zero temperature $T=0$, since the horizon radius $r_h$ remains finite and $\gamma$ independent, the Horndeski coupling is not modifying any of the conductivities; the latter property has been already noticed for the electric conductivity in \cite{Jiang:2017imk}.\\
For completeness we also verified that the Kelvin formula:
\begin{equation}
\frac{\alpha}{\sigma}\Big|_{T=0}\,=\,\lim_{T\rightarrow 0}\,\frac{\partial s}{\partial q}\Big|_T
\end{equation}
introduced in \cite{Davison:2016ngz} holds. This represents another non-trivial test that such a relation could be implied by the presence of an AdS$_2$ near-horizon geometry as suggested in \cite{Blake:2016jnn}.\\
At leading order in the new coupling $\gamma$ we obtain:
\begin{align}
&\sigma\,=\,1\,+\,\frac{q^2}{k^2 \,r_h^2}\,+\,\frac{4 \,\pi\,  \gamma \, q^2\,T}{k^2\,r_h^3}\,+\,\mathcal{O}\left(\gamma^2\right),\\
&\bar{\kappa}\,=\,\frac{16 \,\pi ^2\,T\, r_h^2}{k^2}\,+\,\frac{64\, \pi ^3\, \gamma\,T^2\,r_h}{k^2}\,+\,\mathcal{O}\left(\gamma^2\right),\\&
 \alpha=\bar{\alpha}=\,\frac{4 \,\pi \, q}{k^2}\,+\,\frac{16 \,\pi ^2\, \gamma\,  q\, T}{k^2 \,r_h}\,+\,\mathcal{O}\left(\gamma^2\right),\\
 &\kappa\,=\,\frac{16 \,\pi ^2\, T\, r_h^4}{k^2 \,r_h^2\,+\,q^2}\,+\,\frac{64 \,\pi ^3\, \gamma  \,k^2\, T^2\, r_h^5}{\left(k^2\, r_h^2\,+\,q^2\right)^2}\,+\,\mathcal{O}\left(\gamma^2\right)
\end{align}\\[0.15cm]
which for $\gamma=0$ are in agreement with the known results \cite{Donos:2014cya}.\\
From now on, unless explicitely said, we fix the value of the cosmological constant (in unit of the AdS radius) to be $\Lambda=-3$.\\[0.25cm]
\textbf{Conductivities bounds}\\[0.15cm]
It is interesting to check the behaviour of the previous conductivities in the incoherent limit in order to check the universal bounds proposed in \cite{Grozdanov:2015qia,Grozdanov:2015djs} for \textit{simple} holographic models.\\
\color{black} In an incoherent metal there is no long-lived quantity which overlaps with the current operator $\mathcal{J}$. In particular the momentum operator $\mathcal{P}$, which has a finite overlap with $\mathcal{J}$ as a consequence of a finite charge density $q$, shows a late time behaviour of the form:
\begin{equation}
\langle \mathcal{P}(t) \rangle\,\sim\,e^{-\,t/\tau_{rel}}
\end{equation}
where the relaxation time $\tau_{rel}$ is parametrically smaller than the typical energy scale, \textit{i.e.} $\tau_{rel}\ll k_B T$. As a result the optical response is not well described by the Drude form and indeed no definite Drude peak is present; momentum is quickly dissipated and the physics is dominated by diffusion. Given the inversely proportional relation between the relaxation time $\tau_{rel}$ and the graviton mass $\mathcal{M}$ \eqref{TAU}, from the bulk point of view the incoherent limit can be realized via a parametrically large $\mathcal{M}$.\\
More in details we can define the incoherent limit as:
\begin{equation}\label{INCINC}
\frac{k}{\mu}\,\gg\,1\,,\qquad \frac{k}{T}\,\gg\,1\,,\qquad \textit{with}\,\,\,\,\,\,\,\,\frac{T}{\mu}\,\,\,\,\,\,\,\textit{finite}
\end{equation}
Working at finite charge density $q$, we obtain: 
\begin{align}\label{inco}
&\sigma\,=\,1\,+\,\mathcal{O}\left(1/k^2\right),\\
 &\alpha=\bar{\alpha}=\,\frac{4\,\pi\,q}{k^2},\\
 &\kappa\,=\,\frac{8\,\pi^2\,T}{3}\,+\,\mathcal{O}\left(1/k\right)
\end{align}
\color{black}
As expected from the mass definition \ref{effmass} the Horndeski deformation is not a leading contribution in the incoherent limit and therefore it does not appear in the previous formulae. In other words, the contribution in \ref{effmass} given by the Horndeski coupling $\gamma$ scales in the incoherent limit $\sim k^3$, while the contribution at $\gamma=0$ scales like $\sim k^4$ and it is therefore dominant leading to the same results of the linear Stueckelberg theory.\\[0.15cm]
It is very easy to see from $\eqref{inco}$ that the electric conductivity satisfies:
\begin{equation}
\sigma\,\geq\,1\,\equiv\,\sigma_{bound}
\end{equation}
as proposed in \cite{Grozdanov:2015qia}. It is nowadays clear that in order to affect such a bound one needs to modify the Maxwell term via additional couplings \cite{Baggioli:2016oqk,  Gouteraux:2016wxj, Garcia-Garcia:2016hsd, Baggioli:2016pia} or via non linear deformations \cite{Baggioli:2016oju}.\\
Of more interest is the bound on $\kappa/T$ proposed later in \cite{Grozdanov:2015djs}:
\begin{equation}
\frac{\kappa}{T}\,\geq\,\mathcal{C}
\end{equation}
where $\mathcal{C}$ is a non zero $\mathcal{O}(1)$ number possibly depending on the various parameters and couplings of the model. To the best of our knowledge no counterexamples to such a bound are known yet.\\
In the specific model we considered we obtain:
\begin{align}\label{kappainc}
&\left(\frac{\kappa}{T}\right)^{(inc)}\,=\,\frac{8\,\pi^2}{3}\,+\,\mathcal{O}\left(1/k\right)
\end{align}
meaning that the Horndeski coupling does not affect the universal value of $\kappa/T$ and the bound proposed in \cite{Grozdanov:2015djs} holds. It would be interesting to check this statement in more complicated theories which couple the Stueckelberg scalars to gravity and go beyond the simple toy model considered here. In order to affect such a bound it would be necessary to have a modification of the effective graviton mass \ref{effmass} which scales in the incoherent limit like $\sim k^p$ with $p\geq 4$.\\[0.2cm]
We also investigate the recent bound proposed in \cite{Wu:2017mdl} regarding the incoherent conductivity $\sigma^{\mathcal{I}}$ introduced in \cite{Davison:2015taa}. We consider the incoherent current\footnote{\color{black}In the limit of slow momentum dissipation (in other words, at leading order in its strength) we obtain:
\begin{equation}
\mathcal{J}^{\mathcal{I}}\,\equiv\,\frac{s\,T\,\mathcal{J}\,-\,q\,\mathcal{J}^Q}{s\,T\,+\,\mu\,q}
\end{equation}
 which was analyzed in \cite{Wu:2017mdl}. We thank Shao-Feng Wu for clarifications about this point.\color{black}}:
\begin{equation}\label{ff}
\mathcal{J}^{\mathcal{I}}\,\equiv\,\mathcal{J}\,-\,\frac{\chi_{\mathcal{J}\mathcal{P}}}{\chi_{\mathcal{P}\mathcal{P}}}\,\mathcal{P}
\end{equation}
and its corresponding conductivity $\sigma^{\mathcal{I}}$\footnote{\color{black}Notice that the relation between $\sigma^{\mathcal{I}}$ and the sketchy division made in \eqref{du} is far from trivial and it is well explained in \cite{Davison:2015bea}.\color{black}}\color{black}. By construction such a current has no overlap with the momentum operator $\mathcal{P}$, \textit{i.e.} $\langle\mathcal{J}^{\mathcal{I}}\,\mathcal{P}\rangle\,=\,0$, and it is therefore completely insensitive to the momentum relaxation mechanisms of the system. It represents a purely diffusive mode which could in principle display universal features and saturate universal relations.\\
The authors of \cite{Wu:2017mdl} recently proposed a universal bound for $\sigma^{\mathcal{I}}$ which takes the form:
\begin{equation}\label{BB}
\sigma^{\mathcal{I}}\,\geq\,\left(\sigma\,-\,\alpha^2\,\frac{T}{\bar{\kappa}}\right)\,\frac{\left(s\,T\right)^2}{\left(s\,T\,+\,\mu\,q\right)^2}
\end{equation}
and seems to hold in various holographic theories with momentum dissipation.\\
We verified that the inequality \ref{BB} corresponds in our model to the requirement:
\begin{equation}
\mathcal{M}_h^2\,>\,0
\end{equation}
and it is therefore clearly satisfied at any rate of momentum relaxation\footnote{Notice however that the relation:
\begin{equation}
s\,T\,\alpha\,-\,q\,\bar{\kappa}\,=\,0
\end{equation}
does not hold in our model (and in the ones of \cite{Baggioli:2016pia} neither).}.
\section{Diffusivities and Chaos}\label{sec3}
In this section we are interested in computing the charge and energy diffusivities in the incoherent limit in order to check the universal bounds proposed in \cite{Blake:2016wvh,Blake:2016sud}.\\
\color{black} First we notice that the temperature definition in \eqref{Tdef} implies the equality:
\begin{equation}\label{EEE}
\underbrace{\frac{\gamma}{\pi\,\text{erfi}\left(\frac{\sqrt{\gamma }}{2 \,R_h}\right)^2}\,\left(\frac{\mu}{k}\right)^2\,+\frac{16 \,\pi  \,\rho_h^3\, e^{-\,\frac{\gamma }{4
   \,\rho_h^2}}}{k}\,\frac{T}{k}}\,+\,4\, \Lambda  \,\rho_h^4\,+\,2\, \rho_h^2\,=\,0
\end{equation} 
where we have defined the dimensionless horizon radius $\rho_h=r_h/k$.\\
In the incoherent limit \eqref{INCINC} the underbraced terms in \eqref{EEE} drop to zero and we therefore obtain:
\begin{equation}\label{incdef}
r_h^{(inc)}\,=\,\frac{k}{\sqrt{6}}
\end{equation}
namely the fact that the horizon radius becomes proportional to the strength of momentum dissipation $k$.\color{black}\\
\begin{figure}
\centering
\includegraphics[width=5cm]{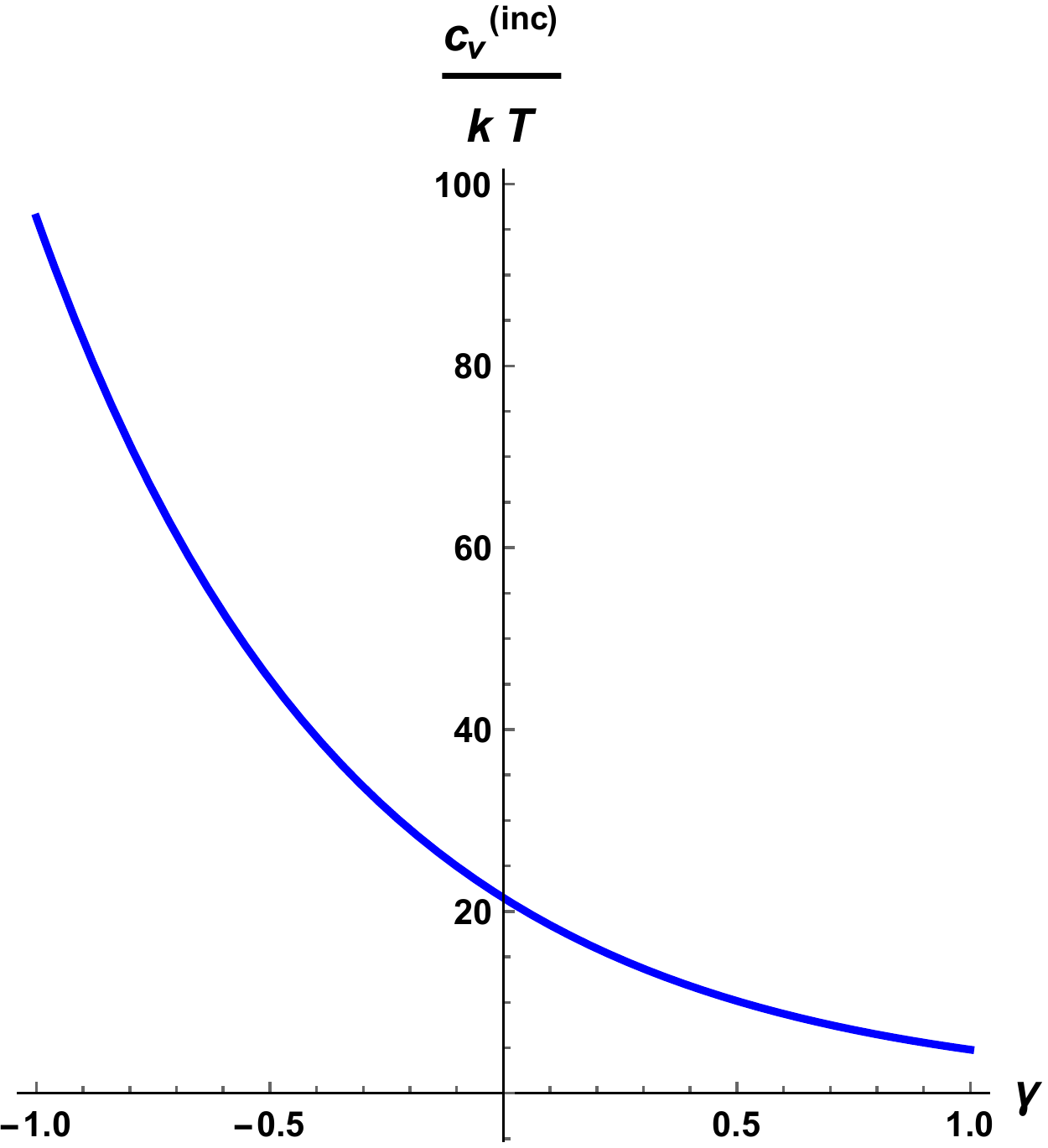}%
\qquad
\includegraphics[width=5cm]{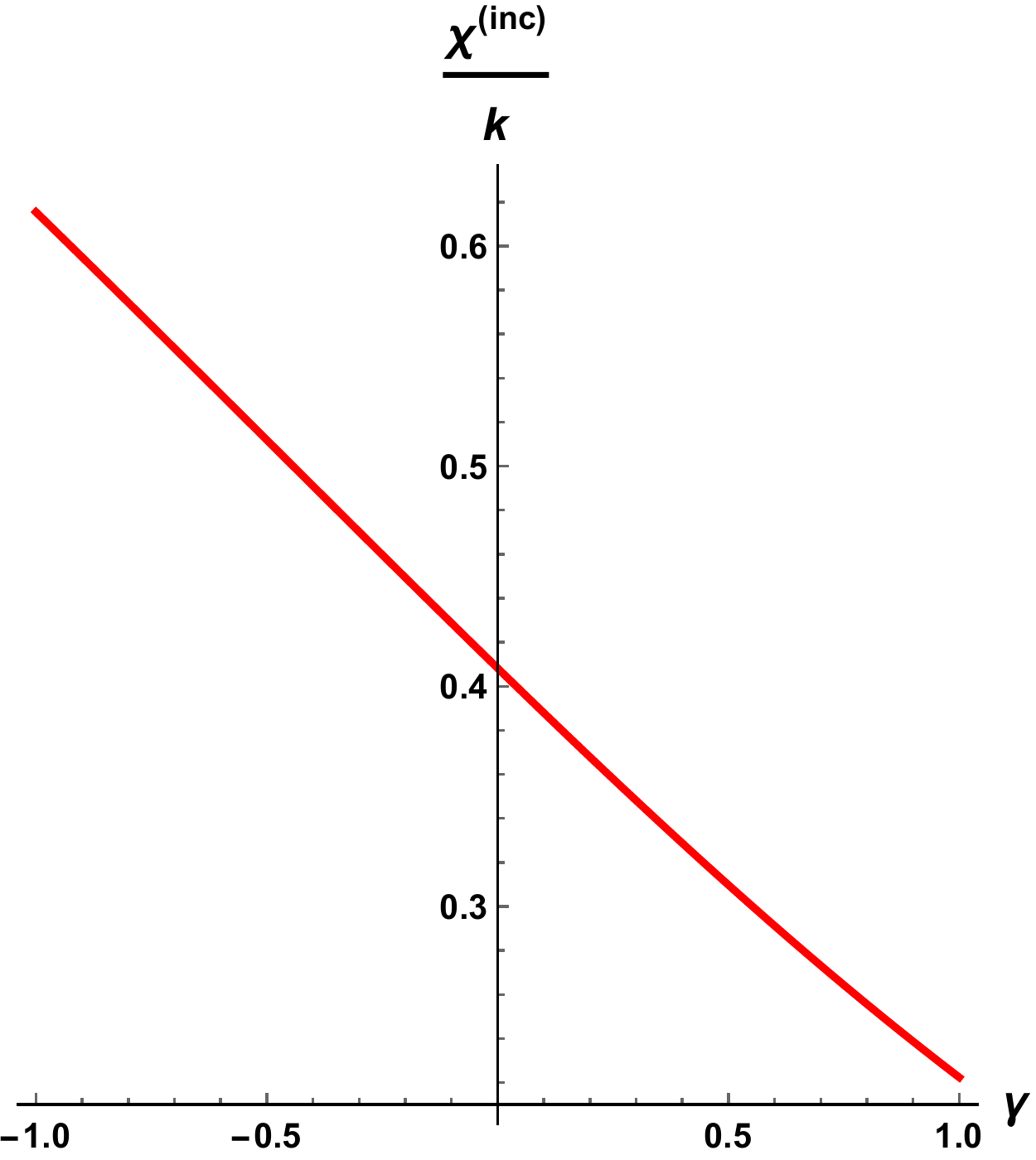}
\caption{Heat capacity and charge susceptibility in the incoherent limit in function of the Horndeski coupling $\gamma$.}
\label{fig1}
\end{figure}
As a first step we compute the heat capacity and the charge susceptibility which are defined as:
\begin{equation}
c_V\,=\,T\,\left(\frac{ds}{dT}\right)_q\,,\qquad \chi\,=\,\left(\frac{\partial q}{\partial \mu}\right)_T
\end{equation}
Given the entropy density \eqref{entropydens} we can derive the heat capacity at constant charge density as:
\begin{equation}
c_V\,=\,T\,\left(\frac{ds}{dr_h}\right)\,\left(\frac{dT}{dr_h}\right)^{-1}\,\Big|_q\,=\,\frac{32 \,\pi ^2\, T \,r_h^5}{\left(k^2 \,r_h^2\,+\,q^2\right) \,e^{\frac{\gamma \, k^2}{4 \,r_h^2}}\,+\,2\, \pi \, T \,r_h\, \left(2 \,r_h^2\,-\,\gamma 
   \,k^2\right)}
  \end{equation}
In the incoherent limit, using \eqref{incdef}, the heat capacity becomes\footnote{Notice that in the incoherent limit the heat capacity at constant charge density coincides with the one at constant chemical potential.} :
\begin{equation}
c_V^{(inc)}\,=\,\frac{8}{3}\, \sqrt{\frac{2}{3}}\, \pi ^2\, e^{-\,3\, \gamma /2}\,k\,T
\end{equation}
which reduces for $\gamma=0$ to the known results.\\
\color{black}In the incoherent regime the horizon radius $r_h$ is independent of the charge \eqref{incdef} and we can directly derive the value of the susceptibility from \eqref{EE3}\color{black}:
\begin{equation}\label{chiINC}
\chi^{(inc)}\,=\,\frac{\sqrt{\gamma }\, k}{\sqrt{\pi }\, \text{erfi}\left(\sqrt{\frac{3\,\gamma}{2} }\right)}\,+\,\mathcal{O}\left(k^0\right)
\end{equation}
We plot the behaviour of the susceptibility and the heat capacity in the incoherent limit in fig.\ref{fig1}. \color{black} The Horndeski coupling does not lead to any important qualitative new feature in the incoherent limit. In the direction of understanding the Horndeski coupling better it would be interesting to study those quantities away from the incoherent limit. One could for example study their temperature and momentum dissipation dependence as done in \cite{Baggioli:2015gsa} for simpler holographic massive gravity models.\color{black}\\
We continue defining the butterfly velocity for our geometry which reads:
\begin{equation}\label{butterfly}
v_B^2\,=\,\frac{\pi\,T\, e^{\frac{\gamma \, k^2}{4 \,r_h^2}}}{r_h}
\end{equation}
The details about its derivation are given in appendix \ref{app2}.\\
In the incoherent limit we obtain:
\begin{equation}
{v_B^2}^{(inc)}\,=\,\frac{\sqrt{6}\, \pi \, e^{3\, \gamma /2}\, T}{k}
\end{equation}
In the incoherent limit charge and energy transport decouple and the corresponding diffusivities can simply be defined as:
\begin{equation}
D_c\,=\,\frac{\sigma}{\chi}\,,\qquad D_e\,=\,\frac{\kappa}{c_V}
\end{equation}
Using all the previous results we finally obtain:
\begin{align}\label{diffu}
\boxed{\color{black}\frac{D_c\,T}{v_B^2}\Big|_{inc}\,=\,\frac{e^{-\,3\, \gamma /2}\, \text{erfi}\left(\sqrt{\frac{3}{2}} \,\sqrt{\gamma }\right)}{\sqrt{6 \,\pi }\, \sqrt{\gamma }}\,,\qquad \frac{D_e\,T}{v_B^2}\Big|_{inc}\,=\,\frac{1}{2\,\pi}}
\end{align}
where the Horndeski coupling takes values inside the range:
\begin{equation}
-\,\infty\,<\,\gamma\,\leq\,\frac{1}{3}
\end{equation}
We plot the results in fig.\ref{fig2}.\\
The results are quite striking. The charge diffusivity acquires a mild dependence on the Horndeski coupling $\gamma$ due to the modification of the static charge susceptibility \ref{chiINC}; despite such a modification we always have:
\begin{equation}
\frac{D_c\,T}{v_B^2}\Big|_{inc}\,\geq\,\mathcal{C}\,>\,0
\end{equation}
as conjectured in \cite{Blake:2016wvh}.\\
The result obtained for the energy diffusion is even more surprising: the Horndeski coupling $\gamma$ does not affect at all the value of $D_eT/v_B^2$ in the incoherent limit, which stays the same of the linear axion theory value \cite{Blake:2016wvh,Blake:2016sud}. \color{black} Notice however that this happens in an highly non trivial way. The butterfly velocity and the heat capacity get both corrected by the Horndeski coupling but in the ratio $D_eT/v_B^2$ the modifications cancel with each other giving the same value of the linear theory.\color{black}\\[0.2cm]
To summarize, the Horndeski coupling $\gamma$ we considered in this work is not providing any interesting violation of the bounds for the diffusivities in the incoherent limit proposed through the butterfly velocity in \cite{Blake:2016wvh,Blake:2016sud}.
We will get back to this point in the conclusions.
\begin{figure}
\centering
\includegraphics[width=7cm]{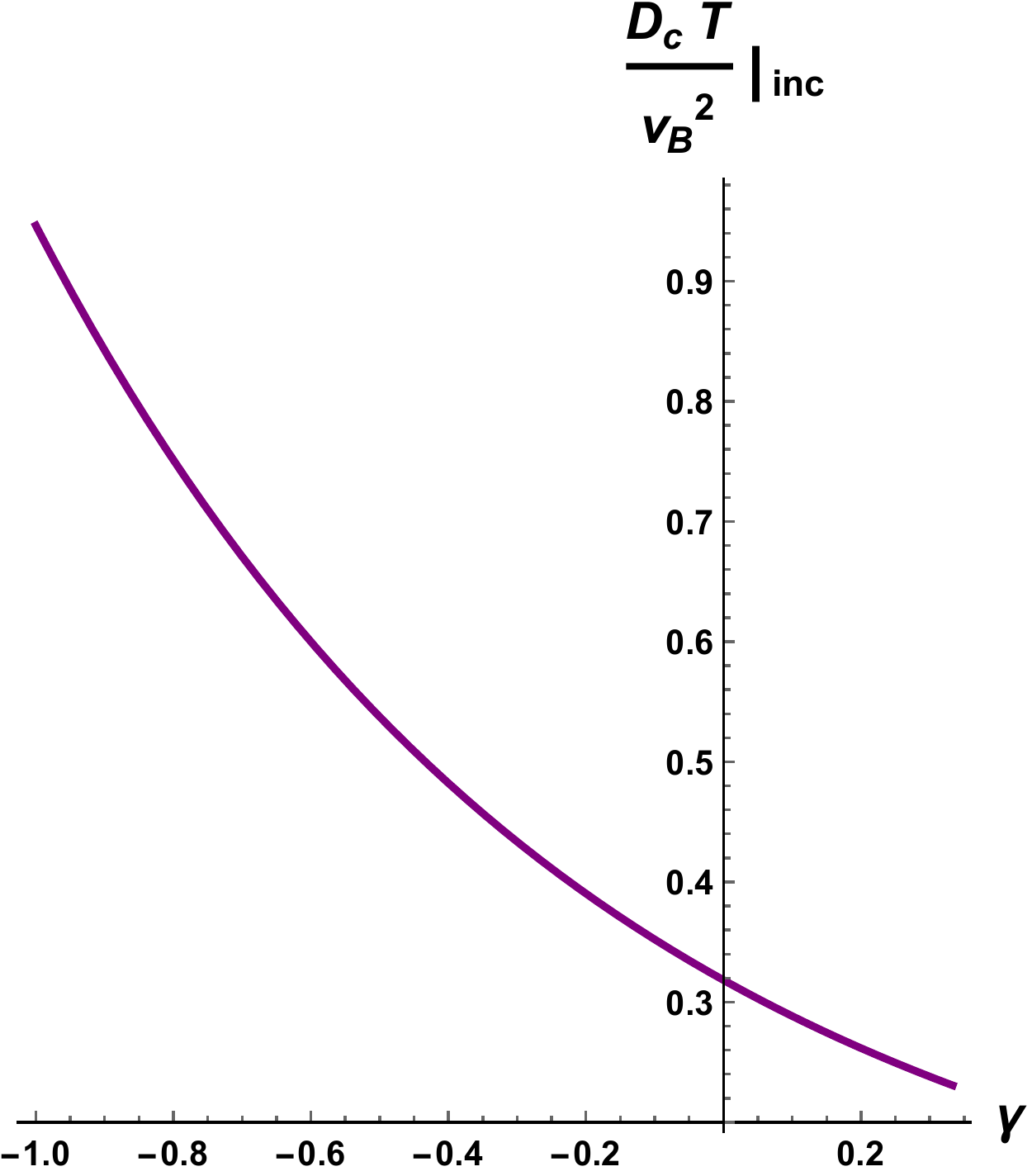}
\caption{Charge diffusivity in the incoherent limit in function of the Horndeski coupling $\gamma$: $\frac{D_c\,T}{v_B^2}\Big|_{inc}$.}
\label{fig2}
\end{figure}
\section{Conclusions}\label{sec4}
The idea of our paper is to check to which extent the universal bounds on heat conductivity and energy diffusivity:
\begin{equation}
\left(\frac{\kappa}{T}\right)^{(inc)}\,\geq\,\mathcal{C}_0\,>\,0\,,\qquad \left(\frac{D_e\,T}{v_B^2}\right)^{(inc)}\,\geq\,\mathcal{C}_1\,>\,0\,.
\end{equation}
proposed in \cite{Grozdanov:2015djs,Blake:2016wvh,Blake:2016sud} are valid in generic holographic bottom up models with momentum dissipation.\\
So far holography has not been able to provide any counterexample. To the best of our knowledge, the only existing violation for the energy diffusion constant bound is derived via a non-homogeneous generalization of the SYK model in \cite{Gu:2017ohj}.\\

In this work we analyzed a recently introduced holographic toy model where the translations breaking sector is directly coupled to the gravitational degrees of freedom through an Horndeski type coupling \cite{Jiang:2017imk}. We find that the Horndeski modification represents a subleading correction in the incoherent limit and it therefore does not modify any of the thermoelectric conductivities. As a consequence the previous universal bounds are not affected and the results we find in the energy sector are exactly identical to the ones of the linear theory \cite{Grozdanov:2015djs,Blake:2016wvh,Blake:2016sud}.\\

It would be certainly interesting to test the same statements in more complicated models like the one proposed in \cite{Garcia-Garcia:2016hsd} containing a coupling of the type:
\begin{equation}
\sim\,\mathcal{H}\left(X\right)\,R\,,\qquad \text{with}\qquad X\,\equiv\,\partial_\mu \phi^I \partial^\mu \phi^I
\end{equation}
it is anyway not clear to us to which extent those represent consistent models where ghosty excitations and gradient instabilities are absent. On the other side it would very promising to find out the holographic gravity dual to the SYK model proposed in \cite{Gu:2017ohj} which could potentially provide a violation of the aforementioned universal bounds.\\

The results of our paper represent another non trivial check that the universal bounds on the energy diffusivity and the thermal conductivity are much more robust than the corresponding ones related to the charge sector. Given the increasing number of positive confirmations, instead of finding counterexamples, it would be very constructive and stimulating to try to prove and understand those statements at least in some simple case, in the direction of \cite{Blake:2016jnn} for example or even outside the framework of holography.\\

As aside interesting questions regarding the model considered in this paper we certainly encounter the behaviour of the $\eta/s$ ratio \cite{futurepaper}, the temperature scalings of the thermoelectric conductivities and the features of the optical AC conductivities.\\[0.3cm]
We hope to return to some of these issues in the near future.
\section*{Aknowledgments}
We would like to thank Elias Kiritsis, Blaise Gouteraux, Andrea Amoretti, Rene Meyer, Mikhail Goykhman, Andy Lucas, Lefteris Papantonopoulos, Kazem Bitaghsir, Keun-Young Kim, Jian-Pin Wu, Mike Blake,  Antonio Garcia Garcia and Aurelio Romero Bermudez for innumerous discussions about the topics presented in this paper. We are particularly grateful to Keun-Young Kim, Andrea Amoretti and Blaise Gouteraux for their valuable comments about the manuscript.\\
We are also grateful to the anonymous referee for his/her helpful and interesting suggestions and comments.\\
MB is supported in part by the Advanced ERC grant SM-grav, No 669288.
WJL is financially supported by  the Fundamental Research Funds for the Central Universities No. DUT 16 RC(3)097 as well as NSFC Grants No. 11375026.
\newpage
\appendix
\section{Details about the entropy density}\label{appen}
The total entropy S of the dual CFT can be expressed as a Noether charge using the Wald formula \cite{Wald:1993nt,Visser:1993nu,Brustein:2007jj}:
\begin{equation}
S\,=\,-2\,\pi\,\int_{\Sigma}\,d^2x\,\sqrt{-h}\,\frac{\delta\,\mathcal{L}}{\delta\,R_{\mu\nu\rho\sigma}}\,\epsilon_{\mu\nu}\,\epsilon_{\rho\sigma}
\end{equation}
where  $\epsilon_{\mu\nu}$ is the binormal on the horizon $\Sigma$ and h the induced metric.\\
For the Einstein-Hilbert action the latter implies the well known area law\footnote{It was actually shown that this formula continues to hold once one replace $G_N$ with the effective gravitational coupling $G_{eff}$ \cite{Brustein:2007jj}.}
\begin{equation}
S\,=\,\frac{\mathcal{A}}{4\,G_N}
\end{equation}
In our case the coupling $\gamma$ could in principle affects the entropy density s. \color{black}We follow the notations given in \cite{Jiang:2017imk}\color{black}.\\
The ansatz \eqref{ansatz} implies that the binormal vector has only two non vanishing components:
\begin{equation}
\epsilon_{tr}\,=\,-\,\epsilon_{rt}\,=\,\sqrt{\frac{h(r)}{f(r)}}\,=\,\sqrt{U(r)}
\end{equation}
and therefore the formula simplifies to:
\begin{equation}
S\,=\,-8\,\pi\,\int_{\Sigma}\,d^2x\,r_h^2\,U(r_h)\,\frac{\delta\,\mathcal{L}}{\delta\,R_{rtrt}}
\end{equation}
which correctly reproduces the Einstein-Hilbert case\footnote{In this case we have $\mathcal{L}\,=\,\frac{R}{16\,\pi\,G_N}$ and $U(r_h)=1$. We obtain: \begin{equation}
\frac{\delta\,\mathcal{L}}{\delta\,R_{rtrt}}\,=\,\frac{1}{32\,\pi\,G_N}\,\left(g^{tt}g^{rr}-g^{tr}g^{tr}\right)
\end{equation}
Since $g^{tr}=0$ and $g^{rr}g^{tt}=-1$ we obtain:
\begin{equation}
S\,=\,\frac{1}{4\,G_N}\,\int_{\Sigma}\,d^2x\,r_h^2\,\,=\,\frac{\mathcal{A}_h}{4\,G_N}
\end{equation} }.\\
Following our action \eqref{themodel} we obtain:
\begin{equation}
\frac{\delta\,\mathcal{L}}{\delta\,R_{\mu\nu\rho\sigma}}\,=\,\underbrace{\frac{1}{2}\,\left(g^{\mu\rho}g^{\nu\sigma}\,-\,g^{\mu\sigma}g^{\nu\rho}\right)}_{EH\,\,action}\,\left(1\,-\,\frac{\gamma}{4}\,g^{\delta \xi}\,X_{\delta \xi}\right)+\,\frac{\gamma}{4}\,\left(X^{\mu\nu}\,g^{\rho\sigma}\,-\,X^{\mu\sigma}\,g^{\rho\nu}\right)
\end{equation}
where we defined $X_{\delta \xi}=\,\sum_i\,\partial_\delta \phi^i\,\partial_\xi \phi^i$.\\
We are interested in the $rtrt$ component which simplifies to:
\begin{equation}
\frac{\delta\,\mathcal{L}}{\delta\,R_{rtrt}}\,=\,\frac{1}{2}\,g^{rr}g^{tt}\,\left(1\,-\,\frac{\gamma}{4}\,g^{\delta \xi}\,X_{\delta \xi}\right)\,=\,-\,\frac{1}{2\,U(r_h)}\,\left(1\,-\,\frac{\gamma}{2\,r_h^2}\,k^2\right)
\end{equation}
We therefore conclude that\footnote{In the language of \cite{Brustein:2007jj} our gravitational effective coupling reads:
\begin{equation}
\frac{1}{G_{eff}}\,=\,\left(1\,-\,\frac{\gamma}{2\,r^2}\,k^2\right)
\end{equation}} :
\begin{equation}
S\,=\,4\,\pi\,\left(1\,-\,\frac{\gamma}{2}\,k^2\right)\int_{\Sigma}\,d^2x\,r_h^2\,\,=\,\left(1\,-\,\frac{\gamma}{2\,r_h^2}\,k^2\right)\,S_{EH}
\end{equation}
Finally the entropy density s for our model reads:
\begin{equation}\label{entropydens}
s\,=\,4\,\pi\,r_h^2\,\left(1\,-\,\frac{\gamma}{2\,r_h^2}\,k^2\right)
\end{equation}
which for $\gamma=0$ reproduces the known results.
\section{Details about the derivation of the DC conductivities}\label{app1}
We follow the generic method introduced in \cite{Donos:2014cya}.\\
We switch on the following set of linear perturbations:
\begin{align}\label{b}
&\delta A_x\,=\,(\zeta \,A_t(r)\,-\,E) \,t\,+\,a_x(r)\,,\nonumber\\
&\delta g_{tx}\,=\,-\,\zeta\, h(r)\, t\,+\,r^2\,h_{tx}(r)\,,\nonumber\\
&\delta g_{rx}\,=\,r^2\,h_{rx}(r)\,,\nonumber\\
&\delta\phi^I\,=\,\psi^x(r)\,.
\end{align}
The linearized equations of motion for such modes, after using the background equations, take the form:
\begin{align}
&f\, \left[2 h^2 \left(A_t'\, a_x'\,+\,r^2\,h_{tx}''+4 r\,h_{tx}'+2 h_{tx}\right)+r\,h\, \left(2 h_{tx} \left(r\,
   {A_t'}^2-r\, h''\,-\,h'\right)-r\, h'\, h_{tx}'\right)+r^2 \,h_{tx}\, {h'}^2\right]+ \nonumber\\&+r\, h\, f'\, \left(h \left(r\, h_{tx}'+2
   h_{tx}\right)-r\, h_{tx}\, h'\right) \,=\,0\,,\label{txeq}\\[0.1cm]
   & r^2 \left[r A_t' (E-\zeta  A_t)+h' \left(\gamma  k f \left(\psi '-k h_{rx}\right)+\zeta  r\right)\right]+h
   \left[k^2 r h_{rx} \left(r^2-\gamma  f\right)+\psi '\left(\gamma  k r f-k r^3\right)+\zeta  \left(-\gamma  k^2-2 r^2\right)\right]\,=\,0\,,\\[0.1cm]
   &h\,f'\, \Big(r^2\, h_{tx}\, A_t'\,+\,h\, a_x'\Big)\,+\,f\, \Big\{-\,r^2\,h_{tx}\, A_t'\, h'\,+\,h\, \Big[2\, r\, \Big(r\,
  h_{tx}\, A_t''\,+\,A_t'\, \Big(r\, h_{tx}'\,+\,2\, h_{tx}\Big)\Big)\,+\,a_x'\, h'\Big]\,+\,2\, h^2\,a_x''
   \Big\}\,=\,0.\label{maxeq}
\end{align}
plus an equation for the $\psi$ mode which is not relevant for the following.\\
From the previous equations we can derive two different radially conserved quantities:
\begin{align}
&\mathcal{J}(r)\,=\,-\,\sqrt{-g}\,F^{rx}\\
&\mathcal{J}^Q(r)\,=\,h^{3/2}\,f^{1/2}\,\left(\frac{\delta g_{tx}}{h}\right)'\,-\,A_t\, \mathcal{J}
\end{align}
which once computed at the boundary $r=\infty$ correspond to the electric and heat currents ${\mathcal{J},\,\mathcal{J}^Q}$. The conservation of the electric current $\mathcal{J}$ follows immediately from the Maxwell equation \eqref{maxeq}, while the conservation of $\mathcal{J}^Q$ can be derived using the tx Einstein equation \eqref{txeq} and the Maxwell equation \eqref{maxeq}; see \cite{Donos:2014cya} for more details.\\
Using the ansatz for the perturbations \eqref{b} we obtain:
\begin{align}
&\mathcal{J}(r)\,=\,-\,\sqrt{f/h}\,\left(h\, a_x'\,+\,A_t'\,r^2\,h_{tx}\right)\\
&\mathcal{J}^Q(r)\,=\,\frac{r\, \sqrt{f}\, h_{tx}\, \left(r\, A_t\, A_t'\,-\,r\, h'\,+\,2\, h\,\right)}{\sqrt{h}}\,+\,A_t\, \sqrt{f\,h} 
   \,a_x'\,+\,r^2\, \sqrt{f\,h} \, h_{tx}'
\end{align}
We want to compute the previous two quantities at the horizon and we have therefore to identify the behaviours of the $A_x$ and $h_{tx},h_{xr}$ fields close to the event horizon $r=r_h$.\\
We obtain the following expressions:
\begin{align}
&a_x'\,=\,-\frac{E}{\sqrt{h\,f}}\,+\,\dots\\
&h_{tx}\,=\,\frac{\sqrt{h} \left(r^2\, f\, h_{rx}\,A_t'\,+\,\zeta\,A_t\right)}{r^2\, \sqrt{f} \,A_t'}
\end{align}
Moreover the xr component of the Einstein equations provides a further constraint which takes the following form:
\begin{equation}
h_{rx}\,=\,\frac{\zeta \, h\, \left(\gamma \, k^2\,+\,2\, r^2\right)\,-\,r^3\, \left(A_t'\, (E\,-\,\zeta\, A_t)\,+\,\zeta\,  h'\right)}{k^2 \,r\, \left(h\,
   \left(r^2\,-\,\gamma\,  f\right)\,-\,\gamma\,  r\, f\, h'\right)}
\end{equation}
We can now compute the charge and heat currents at the horizon $\mathcal{J}(r_h),\,\mathcal{Q}(r_h)$ and we can extract the thermoelectric conductivities as:
\begin{align}
&\sigma\,=\,\frac{\partial \mathcal{J}(r_h)}{\partial E}\Big|_h\,,\qquad \alpha\,=\,\bar{\alpha}\,=\,\frac{\partial \mathcal{J}^Q(r_h)}{\partial E}\Big|_h\,=\,\frac{1}{T}\,\frac{\partial \mathcal{J}(r_h)}{\partial \zeta}\Big|_h\,,\qquad \bar{\kappa}\,=\,\frac{1}{T}\,\frac{\partial \mathcal{J}^Q(r_h)}{\partial \zeta}\Big|_h
\end{align}
where the subscript h indicates that all the quantities have to be computed at the horizon $r=r_h$.\\
Following such a prescription we obtain the formulas shown in the main text \eqref{DC}.
\section{Details about the derivation of the formula for the Butterfly velocity}\label{app2}
For completeness in this section we provide some details regarding the computation of the Butterfly velocity used in the main text. We follow the original papers \cite{Roberts:2014isa,Roberts:2016wdl}.\\
For simplicity we rewrite the Einstein equation in the following form:
\begin{eqnarray}\label{c}
&G_{\mu\nu}-\sum_{i=1}^{2}\frac{\gamma}{2}\left[\frac{1}{2}\partial_\mu \phi^i\partial_\nu \phi^i R-2\partial_\rho \phi^i\partial_{(\mu} \phi^i {R_{\nu)}}^\rho-\partial_\rho \phi^i\partial_\sigma \phi^i {{{R_\mu}^\rho}_\nu}^\sigma+g_{\mu\nu}\partial_\rho \phi^i\partial_\sigma \phi^i R^{\rho\sigma}\,+\,\frac{1}{2}\,G_{\mu\nu}\,\left(\partial \phi^i\right)^2\color{black}\right]\nonumber \\
&\equiv \tilde{G}_{\mu\nu}\,=\,\color{black}\frac{1}{2}\color{black}\,\mathcal{T}_{\mu\nu}
\end{eqnarray}
where the stress tensor $\mathcal{T}_{\mu\nu}$ includes all the terms without derivatives of the metric and $G_{\mu\nu}=R_{\mu\nu}-\frac{1}{2}R g_{\mu\nu}$ is the Einstein tensor\footnote{Notice we fix $16\,\pi\,G_N=1$.}. It is clear from previous works \cite{Roberts:2014isa,Roberts:2016wdl,Blake:2016wvh,Blake:2016sud} that all such a terms are not relevant for determining the butterfly velocity and therefore we can just forget about them.\\
In the holographic picture the butterfly velocity is realized geometrically in terms of a shock-wave propagating in the bulk. Let us consider our geometry in Kruskal coordinates:
\begin{equation}
ds^2\,=\,2\,A(uv)\,du\,dv\,+\,B(uv)\,dx^i\,dx^i
\end{equation}
for which the horizon location $r=r_h$ is mapped into $uv=0$. The Kruskal coordinates are defined:
\begin{eqnarray}\label{d}
u\,v\,=\,-\,e^{\sqrt{f'(r_h)\,h'(r_h)}\,r_*}, \ \ \ u/v\,=\,-\,e^{-\,\sqrt{f'(r_h)\,h'(r_h)}\,t},
\end{eqnarray}
where $dr_*=\frac{dr}{\sqrt{f(r)h(r)}}$. Moreover the functions appearing in the metric are related by the following relations:
\begin{eqnarray}\label{e}
A(uv)\,=\,\,\frac{2}{u\,v}\,\frac{h(r)}{f'(r_h)\,h'(r_h)}, \ \ B(uv)\,=\,r^2
\end{eqnarray}
We perturb the spacetime with an operator at $x=0$ and $t_L=t_W$, \textit{i.e.} a localized shock-wave; the butterfly velocity corresponds to the rate of growth of this perturbation.\\
The localized stress tensor of such a perturbation is given by:
\begin{equation}
T_{uu}^{shock}\,=\,E_0\,e^{2\,\pi\,t_W/\beta}\,\delta(u)\,\delta(x)
\end{equation}
where $\beta=\frac{1}{T}$. The shock-wave corresponds to a solution where there is a shift $v\rightarrow v\,+\,h(x,t_W)$ once one crosses the horizon $u=0$. The backreaction produces a perturbation in the spacetime metric of the form:
\begin{equation}
ds^2\,=\,2\,A(uv)\,du\,dv\,+\,B(uv)\,dx^i\,dx^i\,-\,2\,A(uv)\,h(x,t_W)\,\delta(u)\,du^2
\end{equation}
and the stress tensor gets modified accordingly as:
\begin{equation}
\delta \mathcal{T}_{uu}\,=\,T_{uu}^{shock}\,+\,\delta g_{uu}\,g^{\mu\nu}\,T_{\mu\nu}^0\,=\,E_0\,e^{2\,\pi\,t_W/\beta}\,\delta(u)\,\delta(x)\,-\,\color{black}4\color{black}\,h(x,t_W)\,\delta(u)\,\tilde{G}_{uv}^0
\end{equation}
where:
\begin{equation}
\tilde{G}^0_{uv}=G^0_{uv}-\sum_{i=1}^{2}\frac{\gamma}{2}\left[-\partial_\rho \phi^i\partial_\sigma \phi^i {{{R^0_u}^\rho}_v}^\sigma+g^0_{uv}\partial_\rho \phi^i\partial_\sigma \phi^i R^{0\rho\sigma}\,+\,\frac{1}{2}\,G_{uv}^0\,\left(\partial \phi^i\right)^2\color{black}\right]
\end{equation}
The only left and relevant Einstein equation is:
\begin{equation}
G^1_{uu}-\sum_{i=1}^{2}\frac{\gamma}{2}\left[-\partial_\rho \phi^i\partial_\sigma \phi^i {{{R^1_u}^\rho}_u}^\sigma+g^1_{uu}\partial_\rho \phi^i\partial_\sigma \phi^i R^{0\rho\sigma}\,+\,\frac{1}{2}\,G_{uu}^1\,\left(\partial \phi^i\right)^2\color{black}\right]=\color{black}\frac{1}{2}\color{black}\delta \mathcal{T}^1_{uu}
\end{equation}
Setting $\rho=\sigma=i$ we are left with:
\begin{eqnarray}\label{c2}
\left(1\,-\,\frac
{\gamma\,k^2}{2}\,g^{ii}_0\right)\,G^1_{uu}+\gamma k^2\left( {{{R^1_u}^i}_u}^i-g^1_{uu}R^{0ii}\right)=\color{black}\frac{1}{2}\color{black}T^1_{uu},\\
\tilde{G}^0_{uv}=\left(1\,-\,\frac
{\gamma\,k^2}{2}\,g^{ii}_0\right)\,G^0_{uv}+\gamma k^2\left[ {{{R^0_u}^i}_v}^i-g^0_{uv} R^{0ii}\right]
\end{eqnarray}
where the subfix $_0$ refers to the background quantities and the subfix $_1$ refers to the linearly perturbed quantities.\\
From the Einstein equations we can derive\footnote{Notice the identity $u\,\delta'(u)\,=\,-\,\delta(u)$.} the linearized dynamics for the shift $h(x,t_W)$ which takes the form:
\begin{eqnarray}\label{c4}
\left[\partial_i^2-m^2\right]h(x^i,t_w)=\frac{B(0)\,E_0 e^{2\pi t_w/\beta}\delta(x)}{\color{black}2\color{black}\,A(0)}
\end{eqnarray}
where the effective mass reads:
\begin{eqnarray}\label{c5}
m^2=\frac{B'(0)}{A(0)}
\end{eqnarray}
Its expression is surprisingly not modified by the $\gamma$ coupling. Notice that generically higher derivative corrections in the gravity sector might modify such a definition \cite{Alishahiha:2016cjk}.\\
Solving the previous equation we find out that at large distances the solution takes the form:
\begin{equation}\label{OTOC1}
h(x,t_W)\,\sim\,\frac{E_0\,e^{\frac{2\,\pi}{\beta}(t_W\,-\,t^*)\,-\,m\,|x|}}{|x|^{1/2}}
\end{equation}
\color{black}where $t^*=\frac{\beta}{2\,\pi}\log \frac{1}{G_N}$ is the scrambling time \cite{Roberts:2016wdl}.\\ The null shift along the $v$ direction parametrized by the bulk solution $h(x,t_W)$ corresponds (in the dual side) to the commutator of the operator $\mathcal{W}$ inserted at different times $t=[0,t_W]$; in other words the solution \eqref{OTOC1} is the bulk reincarnation of the exponential behavior in \eqref{OTOC}. We refer to \cite{Shenker:2013yza,Roberts:2014isa} for further details. \color{black}\\
Then, from \eqref{OTOC1} we can deduce the Lyapunov exponent and the butterfly velocity:
\begin{equation}
\lambda_L\,=\,\frac{2\,\pi}{\beta}\,,\qquad v_B\,=\,\frac{2\,\pi}{\beta\,m}
\end{equation}
where $\beta=1/T$.\\
The final step is to re-express $A(0)$, $B(0)$ and $B'(0)$  in the original $(t,r,x^i)$ coordinates. 
Near the horizon we have:
\begin{eqnarray}\label{d}
e^{\sqrt{f'(r_h)h'(r_h)}r_*}\approx \kappa_0\, (r-r_h)+...\\
h(r)\approx h'(r_h) \,(r-r_h)+...
\end{eqnarray}
therefore we obtain:
\begin{eqnarray}\label{e}
&&A(0)\approx \,-\,\frac{2}{\kappa_0\, f'(r_h)}+...\\
&&B'(0)\approx \,-\,\frac{2 \,r_h}{\kappa_0}+..., 
\end{eqnarray}
where $\kappa_0$ is a positive constant and finally:
\begin{equation}
m^2\,=\,r_h\,f'(r_h)
\end{equation}
All in all the butterfly velocity, expressed in the original coordinates, takes the form:
\begin{equation}
v_B^2\,=\,\frac{\pi\,T\, e^{\frac{\gamma \, k^2}{4 \,r_h^2}}}{r_h}
\end{equation}
which is the expression \ref{butterfly} appearing in the main text and it clearly agrees with the results for $\gamma=0$ \cite{Blake:2016wvh}.
\bibliographystyle{JHEP}
\bibliography{popebib}
\end{document}